\documentstyle[aps,multicol,pra,amsfonts]{revtex}
\tighten
\newtheorem{Def}{Definition}
\newtheorem{Lemma}{Lemma}
\newtheorem{Thrm}{Theorem}
\newtheorem{Cor}{Corollary}
\newtheorem{Eg}{Example}
\begin{document}
\draft
\title{Good Quantum Convolutional Error Correction Codes And Their Decoding
 Algorithm Exist}
\author{H. F. Chau\footnote{e-mail: hfchau@hkusua.hku.hk}}
\address{Department of Physics, University of Hong Kong, Pokfulam Road, Hong
 Kong}
\date{\today}
\preprint{quant-ph/9806032}
\maketitle
\begin{abstract}
 Quantum convolutional code was introduced recently as an alternative way to
 protect vital quantum information. To complete the analysis of quantum
 convolutional code, I report a way to decode certain quantum convolutional
 codes based on the classical Viterbi decoding algorithm. This decoding
 algorithm is optimal for a memoryless channel. I also report three simple
 criteria to test if decoding errors in a quantum convolutional code will
 terminate after a finite number of decoding steps whenever the Hilbert space
 dimension of each quantum register is a prime power. Finally, I show that
 certain quantum convolutional codes are in fact stabilizer codes. And hence,
 these quantum stabilizer convolutional codes have fault-tolerant
 implementations.
\end{abstract}
\par\medskip\noindent
\pacs{PACS numbers: 03.67.-a, 03.67.Dd, 89.70.+c, 89.80.+h}
\begin{multicols}{2}
\narrowtext
\section{Introduction}
\label{S:Intro}
 Quantum error correcting codes (QECCs) and their fault tolerant
 implementations are effective ways to protect and to manipulate quantum
 information in the presence of noise. A QECC works by adding suitable
 redundancy in the form of entanglement to the original quantum state in such a
 way that one can reconstruct the original state after decoherence. Since the
 discovery of QECC by Shor \cite{9-bit}, researchers have discovered many ways
 to construct QECCs. (See, for example,
 Refs.~\cite{5-bit,St1,Cald3,Cald2,Gott1,9-reg,5-reg,N-add,AnE1,AnE2,Cond2}.)
 These QECCs share a common characteristic, namely, one first divides the
 original quantum registers into separate blocks of a fixed finite length. One
 then applies the same encoding method to quantum registers in each block.
 Hence, this kind of codes are called quantum block codes (QBCs). A QECC can be
 decoded by first measuring the error syndromes of the encoded quantum state,
 and then by applying a necessary unitary transformation to the corresponding
 erroneous quantum registers \cite{9-bit,Gott1}. For QBCs, this can be carried
 out in a block by block basis. Since there is only a finite number of error
 syndromes and hence also a finite number of recovery operations in each block,
 decoding a QBC requires only a finite amount of work per (decoded) quantum
 register.
\par
 Recently, Chau constructed another class of codes, known as quantum
 convolutional codes (QCCs), whose encoding scheme for each block depends on
 the states of several other blocks \cite{QCC}. For example, he showed that the
 QCC
\begin{eqnarray}
 & & |k_1,k_2,\ldots \rangle \longmapsto |{\bf k}_{\rm L} \rangle \equiv
 \nonumber \\ & & \bigotimes_{i=1}^{+\infty} \left[ \sum_{p_1,q_1,\ldots}
 \!\frac{1}{N} \,\omega_N^{(k_i + k_{i-2}) p_i + (k_i + k_{i-1} + k_{i-2})
 q_i} \,|p_i\!+\!p_{i-1}, \right. \nonumber \\ & & ~~~~~\left.
 \raisebox{1.8em}{~} p_i\!+\!p_{i-1}\!+\!q_{i-1},q_i\!+\!q_{i-1},
 q_i\!+\!q_{i-1}\!+\!p_i \rangle \right] ~, \label{E:14QCC}
\end{eqnarray}
 where $k_i \in {\Bbb Z}_N$ for all $i>0$, $k_j = 0$ for all $j\leq 0$,
 $\omega_N$ is a primitive $N$th root of unity, the sum is from 0 to $N\!-\!1$,
 and all additions in the state ket are modulo $N$, is capable of correcting
 one error out of every eight consecutive quantum registers.
\par
 While QCCs are of interest of their own right, it is not clear how to decode
 them. This is because the length of the original quantum state and hence the
 number of correctable errors by the code may both be infinite. Furthermore,
 decoding errors may propagate from one block to another due to their
 convolutional nature. Besides, it is not obvious how to manipulate a QCC
 fault-tolerantly.
\par
 In this Paper, I address the questions of decoding and fault-tolerantly
 manipulating a QCC as well as a condition for the existence of a QCC whose
 decoding error does not propagate indefinitely. My key observation is that
 many classical convolutional code as well as quantum block code concepts can
 be extended to the quantum case when one performs the relevant operations
 carefully. I first show that the well-known Viterbi decoding algorithm (VDA)
 \cite{Viterbi,Viterbi_Review} for classical convolutional codes can be
 generalized to QCCs. Then, I show that the quantum version of Viterbi decoding
 (QVA) is equivalent to the maximum likelihood decoding. And hence, QVA is
 optimal in a memoryless channel. After that, I investigate the decoding error
 propagation in QCCs. In particular, I prove three equivalent criteria for QCCs
 to have finite decoding error propagation whenever the Hilbert space dimension
 of each quantum register is a prime power. And finally, I address the question
 of fault-tolerant manipulation of QCCs by showing that the well-known
 fault-tolerant stabilizer code theory can be generalized to QCC.
\section{Classical And Quantum Viterbi Decoding Algorithms}
\label{S:Viterbi}
 Before I go on, let me stress that in my subsequent discussion, I shall
 restrict myself to consider those classical or quantum convolutional codes
 whose encoding can be implemented by a $k$-input $n$-output (and hence also
 $n\!-\!k$ preset registers in the quantum case) $m$-memory (that is, the
 encoding scheme depends on the state of the previous $m$ blocks) quantum
 sequential circuit. (Compare with the definition of classical convolutional
 codes in Ref.~\cite{Forney}.) Actually, all useful convolutional codes belong
 to this category.
\par
 Now, let me begin by briefly reviewing VDA for binary signals
 \cite{Viterbi,Viterbi_Review,Forney,Convolution_Book}. The algorithm starts by
 computing the Hamming distances between the first $d n$ bits of the signal in
 the encoded sequence with the first $d n$ bits of the $2^{d k}$ possible code
 branches where $d = \left\lfloor \frac{k m}{n-k}\!+\!1\right\rfloor$. One then
 keeps only those $2^{(d-1)k}$ code branches with small Hamming distances. (In
 case of a tie in Hamming distances, one keeps the corresponding code branches
 arbitrarily.) Now, one computes the Hamming distances between the first
 $(d\!+\!1) n$ bits of signal with the first $(d\!+\!1)n$ bits of all possible
 code branches that are consistent with a previously kept code branch. This
 process is repeated until either the signal ends for signals of finite length
 or the process is repeated definitely for an infinitely long signal. The final
 surviving code branch is the decoded signal. In essence, VDA tries to find a
 codeword with the smallest Hamming distance from the signal
 \cite{Viterbi_Review,Forney,Convolution_Book,Omura,Forney2,Viterbi2,Fano}.
\par
 Clearly, VDA in the above form cannot be applied directly to QCCs as it
 requires a complete knowledge of the encoded signal. However, if one examines
 the algorithm carefully, it becomes clear that what really required are the
 {\em error syndromes} of the encoded signal for different possible code
 branches. Consequently, VDA can be applied to quantum signals. Suppose
 ${\cal R}$ spans of the set of all error recovery (unitary) operators for
 every $n$ consecutive quantum registers. (For a general error correcting code,
 ${\cal R}$ can be chosen to be in the form $\bigotimes_{i=1}^k\left( f_i \circ
 s_i \right)$ where $f_i$ is either an identity or a controlled phase shift
 operator on a single quantum register, and $s_i$ is either an identity or a
 spin permutation operator on a single quantum register. Consequently, there
 are $\left( N\,N! \right)^k$ elements in ${\cal R}$.) Recall that error
 syndromes can be regarded as operators whose actions have no effect on an
 error-free encoded quantum state. (For instance, the action of error syndrome
 on a stabilizer code simply permutes the stabilizer.) Thus, by measuring
 the eigenvalue of an error syndrome, one gains some information on the
 location and kind of error occurred in a quantum code \cite{St1,Gott1}. Since
 the set of all error syndromes is closed under composition and that the QCC is
 of finite memory, there are only a finite number of independent error syndrome
 operators that acts only on the first $d$ blocks of encoded quantum registers.
 So, for each $R\in {\cal R}^{\otimes d}$, I use a finite (and fixed) number of
 ancillary quantum registers to measure the error syndromes of the first $d n$
 quantum registers after subjecting to the unitary operation $R$. In this way,
 I can the erroneous registers for each $R\in {\cal R}^{\otimes d}$.
\par
 Once the erroneous quantum registers are located, how are we going to correct
 the quantum errors? To answer this question, I have to introduce the following
 definition first.
\begin{Def}
 Let $|\Psi_1\rangle$ and $|\Psi_2\rangle$ be two quantum signals of possibly
 infinite lengths. If it is not possible to find a unitary operator involving a
 finite number of quantum registers which maps $|\Psi_1\rangle$ to $|\Psi_2
 \rangle$, then I say that the {\bf quantum Hamming distance} between $|\Psi_1
 \rangle$ and $|\Psi_2\rangle$ is infinite. Otherwise, I define the
 {\bf quantum Hamming distance} between these two quantum signals as the
 minimum number of quantum registers involved in unitarily transforming from
 one state to the other.
 \par
 Similarly, I define the minimum quantum Hamming distance between a quantum
 signal $|\Psi\rangle$ and the set of all possible codewords of a QECC $C$ to
 be infinite if the quantum Hamming distances between $|\Psi\rangle$ and all
 codewords of $C$ are infinite. Otherwise, I define the quantum Hamming
 distance between $|\Psi\rangle$ and $C$ to be the minimum possible quantum
 Hamming distance between $|\Psi\rangle$ and the codewords in $C$. And for
 simplicity, I shall simply call the minimum quantum Hamming distance between
 $|\Psi\rangle$ and $C$ the {\bf quantum Hamming distance} of $|\Psi\rangle$.
 \par
 I also define the {\bf recovery cost} of bringing the first $d n$ quantum
 signals with respected to $R\in {\cal R}^{\otimes d}$ to be the quantum
 Hamming distance of the quantum signal plus the (minimum) number of registers
 affected by $R$. \label{Def:QHD}
\end{Def}
\par\indent
 Readers can easily check that quantum Hamming distance between two quantum
 signals is a metric for the set of all quantum signals. Moreover, in a loose
 sense, the recovery cost measures how close and how much work is required to
 bring a quantum signal to a quantum codeword.
\par\indent
 With the above definition, I am ready to report QVA decoding: By carefully
 measuring the error syndromes using ancillary quantum registers, I compute the
 recovery cost for the first $d n$ quantum registers for each $R\in
 {\cal R}^{\otimes d}$. I keep $|{\cal R}|^{d-1}$ error recovery operators with
 small recovery costs out of the $|{\cal R}|^d$ possible ones (where
 $|{\cal R}|$ denotes the number of elements in the set ${\cal R}$). Then, I go
 on to compute the recovery cost for the first $(d\!+\!1)n$ quantum signals
 with respected to the set of all possible recovery operators in
 ${\cal R}^{\otimes (d+1)}$ that are consistent with a previously kept recovery
 operator in ${\cal R}^{\otimes d}$. For a quantum signal of finite length, I
 repeat this process until the quantum signal terminates. Then, I regard the
 error of the signal to be caused by the one that produces the minimum possible
 recovery cost among those $|{\cal R}|^d$ ones I kept at the end. And I correct
 the quantum signal accordingly. For quantum signals of infinite length, I have
 to repeat the recovery cost selection process infinitely many times in order
 to find the minimum recovery cost path. In practice, we usually run the QVA
 over a large but finite number of quantum blocks and decode the signal in each
 block separately. The length of such quantum block is usually adaptive; that
 is to say, it is chosen in such a way that the recovery cost paths retained do
 not differ very much from each other. In this way, the effect due to the
 choice of the length of the block is minimized. In summary, regardless of the
 length of the quantum signal, QVA uses a finite number of operations on
 average to recover it.
\par
 Finally, I need to convert the recovered encoded signal to its unencoded form.
 Since I may have an infinitely long signal, the usual trick of running the
 reversible encoding quantum circuit backward does not work. Hence, I have no
 choice but to decode the signal starting from the first encoded block.
 Remember that by including the preset quantum registers, the encoding process
 can be represented by a unitary transformation. Let $C$ is a $k$-input
 $n$-output $m$-memory QCC and $|\Psi_1\rangle$ be a quantum signal (with
 preset quantum registers added). I denote the encoding process for this
 quantum code $C$ and quantum signal $|\Psi\rangle$ by $U_1$. Moreover, I
 denote the encoding process by the same code $C$ on the quantum signal
 $|\Psi_i\rangle$ by $U_i$ where $|\Psi_i\rangle$ equals $|\Psi_1\rangle$
 except that the first $k(i\!-\!1)$ quantum registers are set to zero. Using
 this notation, I write the encoding process $U_1$ as $( U_1 \circ U_2^{-1} )
 \circ ( U_2 \circ U_3^{-1} ) \circ \cdots $. Consequently, the decoding
 process is given by
\begin{eqnarray}
 U_1^{-1} & = & \cdots \circ \left( U_3 \circ U_4^{-1} \right)^{-1} \!\!\circ
 \left( U_2 \circ U_3^{-1} \right)^{-1} \!\!\circ \left( U_1 \circ U_2^{-1}
 \right)^{-1} \nonumber \\ & = & \cdots \circ \left( U_4 \circ U_3^{-1} \right)
 \circ \left( U_3 \circ U_2^{-1} \right) \circ \left( U_2 \circ U_1^{-1}
 \right) ~. \label{E:Gen_Decode}
\end{eqnarray}
 Note that for the $m$-memory code $C$, $U_{i+1} \circ U_i^{-1}$ is a unitary
 operator acting only on the $i$th, $(i\!+\!1)$th, up to $(i\!+\!m)$th encoded
 blocks for all $i$. Moreover, it is easy to check that the action of $U_2
 \circ U_1^{-1}$ is to extract the state of the first unencoded block of
 quantum registers out of the encoded state. After that, the action of $U_3
 \circ U_2^{-1}$ is to extract the state of the second unencoded block out of
 the encoded state, and so on. Thus, Eq.~(\ref{E:Gen_Decode}) gives us a way to
 decode the QCC $C$.
\begin{Eg}
 It is easy to compute $U_{i+1} \circ U_i^{-1}$ in practice. For example, to
 decode the QCC in Eq.~(\ref{E:14QCC}) in Eq.~(\ref{E:Ex_Decode_1}), I
 subtracts the second quantum register by the first, the fourth by the first
 and the third, the fifth by the first, the sixth by the first and the fourth,
 the seventh by the third, and finally the eighth by the third. The resultant
 quantum state is given by Eq.~(\ref{E:Ex_Decode_2}). Then, I discrete inverse
 Fourier transform the first quantum register. The resultant state is given by
 Eq.~(\ref{E:Ex_Decode_3}) after summing over the dummy index $p_1$. Finally,
 it is straight forward to unitarily convert the state in
 Eq.~(\ref{E:Ex_Decode_3}) to the state in Eq.~(\ref{E:Ex_Decode_4}) by
 multiplying a phase proportional to the product of the first quantum register
 by the sum of the third and the twelve registers, and then followed by
 discrete Fourier transforming the third quantum register. The result of all
 these steps above gives $U_2 \circ U_1^{-1}$.
 \begin{mathletters}
 \begin{eqnarray}
  & & \sum_{p_1,q_1,\ldots} \frac{1}{N} \omega_N^{k_1 (p_1+q_1)} |p_1,p_1,q_1,
  p_1\!+\!q_1\rangle \nonumber \\ & & ~~\otimes \frac{1}{N} \omega_N^{k_2 p_2 +
  (k_2+k_1) q_2} |p_1\!+\!p_2,p_1\!+\!p_2\!+\!q_1,q_2\!+\!q_1\rangle \nonumber
  \\ & & ~~\otimes |p_2\!+\!q_2\!+\!q_1\rangle \otimes \frac{1}{N}
  \omega_N^{(k_3+k_1)p_3+(k_3+k_2+k_1)q_3} \nonumber \\ & & ~~\times
  |p_3\!+\!p_2,p_3\!+\!p_2\!+\!q_2,q_3\!+\!q_2\rangle \nonumber \\ & &
  ~~\otimes |p_3\!+\!q_3\!+\!q_2 \rangle \otimes \cdots
  \label{E:Ex_Decode_1} \\
  & \mapsto & \sum_{p_1,q_1,\ldots} \frac{1}{N} \omega_N^{k_1 (p_1+q_1)} |p_1,
  0,q_1,0\rangle \otimes \frac{1}{N} \omega_N^{k_2 p_2 + (k_2 + k_1) q_2}
  \nonumber \\ & & ~~\times |p_2,p_2,q_2,p_2\!+\!q_2\rangle \otimes \frac{1}{N}
  \omega_N^{(k_3+k_1)p_3 + (k_3+k_2+k_1) q_3} \nonumber \\ & & ~~\times
  |p_3\!+\!p_2,p_3\!+\!p_2\!+\!q_3,q_3\!+\!q_2\rangle \nonumber \\ & &
  ~~\otimes |p_3\!+\!q_3\!+\!q_2\rangle \otimes \cdots
  \label{E:Ex_Decode_2} \\
  & \mapsto & \sum_{\lambda,p_1,q_1,\ldots} \frac{1}{N^{3/2}} \omega_N^{k_1
  (p_1+q_1) - p_1 \lambda} |\lambda,0,q_1,0\rangle \nonumber \\ & & ~~\otimes
  \frac{1}{N} \omega_N^{k_2 p_2 + (k_2+k_1) q_2} |p_2,p_2,q_2,p_2\!+\!q_2
  \rangle \nonumber \\ & & ~~\otimes \frac{1}{N} \omega_N^{(k_3+k_1) p_3 +
  (k_3+k_2+k_1) q_3} |p_3\!+\!p_2,p_3\!+\!p_2\!+\!q_3\rangle \nonumber \\ & &
  ~~\otimes |q_3\!+\!q_2,p_3\!+\!q_3\!+\!q_2\rangle \otimes \cdots
  \nonumber \\
  & = & \sum_{q_1,p_2,q_2,\ldots} \frac{1}{\sqrt{N}} \omega_N^{k_1 q_1} |k_1,0,
  q_1,0\rangle \otimes \frac{1}{N} \omega_N^{k_2 p_2 + (k_2+k_1) q_2} \nonumber
  \\ & & ~~\times |p_2,p_2,q_2,p_2\!+\!q_2\rangle \otimes \frac{1}{N}
  \omega_N^{(k_3+k_1) p_3 + (k_3+k_2+k_1) q_3} \nonumber \\ & & ~~\times
  |p_3\!+\!p_2,p_3\!+\!p_2\!+\!q_3,q_3\!+\!q_2\rangle \nonumber \\ & &
  ~~\otimes |p_3\!+\!q_3\!+\!q_2\rangle \otimes \cdots
  \label{E:Ex_Decode_3} \\
  & \mapsto & \sum_{p_2,q_2,\ldots} |k_1,0,0,0\rangle \otimes \frac{1}{N}
  \omega_N^{k_2 (p_2+q_2)} |p_2,p_2,q_2,p_2\!+\!q_2\rangle \nonumber \\ & &
  ~~\otimes \frac{1}{N} \omega_N^{k_3 p_3 + (k_3+k_2) q_3} |p_3\!+\!p_2,
  p_3\!+\!p_2\!+\!q_2,q_3\!+\!q_2\rangle \nonumber \\ & & ~~\otimes
  |p_3\!+\!q_3\!+\!q_2 \rangle \otimes \cdots
  \label{E:Ex_Decode_4}
 \end{eqnarray}
 \end{mathletters}
 In this way, I decode the first quantum register from the code using a finite
 number of two body operators. And inductively, I can decode the rest of the
 encoded quantum signal efficiently. \label{Eg:Decoding}
\end{Eg}
\par\indent
 Now, I move on to prove the optimality of QVA for a memoryless quantum
 channel, namely, a noisy quantum channel whose error occurs randomly and
 independently. Similar to VDA, QVA tries to search for a solution in the
 codeword space with a minimum recovery cost from the signal. (I choose to
 minimize the recovery cost instead of simply the quantum Hamming distance
 because certain fault tolerant operations $U$ for the QCC may belong to
 $\bigotimes_{i=1}^{+\infty} {\cal R}$. In this case, the quantum Hamming
 distances of $|\Psi\rangle$ and $U|\Psi\rangle$ agree although they have
 different recovery costs.) After transmitting an arbitrary unknown encoded
 quantum state through a memoryless channel, the effect of decoherence can be
 regarded as a Markov process. In other words, the probability that the error
 recovery operator needed to act on the $(t\!+\!1)$th block of quantum signals,
 given all error recovery operators for the first $t$ blocks, depends only on
 that of the $(t\!+\!1)$th block. Since I do not manipulate my encoded quantum
 signal during its transmission, I can always conceptually regard my error
 syndrome measurement to be performed immediately after the errors are
 introduced into the signal. But once I have measured the error syndromes,
 the location as well as the type of error each quantum register is suffering
 becomes classical data. Therefore, the effect of a quantum memoryless channel
 is the same as that in a classical probabilistic memoryless channel. More
 precisely, I can always model the chance for a certain quantum error $R\in
 {\cal R}$ to error in a quantum register by a classical probability function.
 (Compare to the argument used in the proof of the security of quantum key
 distribution in Ref.~\cite{QKD_Sec}.) Consequently, the optimality proof of
 VDA \cite{Viterbi_Review,Forney,Convolution_Book,Omura,Forney2} carries over
 directly to QVA.
\par
 Similar to classical convolutional codes, there are two important
 probabilities which measure the performance of a QCC. The first one is called
 the error probability $P_e (E)$, which is defined to be the probability that
 a wrong decoding path is chosen at any given timestep. And the second one is
 called the qubit error probability $P_b (E)$, which is defined to be the
 expected number of information qubit decoding errors per decoded information
 qubit. For $N = 2$ in a binary symmetric channel, these two probabilities are
 given by \cite{Convolution_Book}
\begin{mathletters}
\begin{equation}
 P_e (E) \lesssim A_d 2^d p^{t / 2} ~, \label{E:P_e}
\end{equation}
 and
\begin{equation}
 P_b (E) \lesssim \frac{B_d}{k} 2^d p^{d / 2} ~, \label{E:P_b}
\end{equation}
\end{mathletters}
 where $d$ is the minimum quantum Hamming distance between codewords, $A_d$ is
 the number of mutually orthogonal encoded states of quantum Hamming weight
 $d$, $B_d$ is the number of non-zero mutually orthogonal information qubits on
 all weight $d$ paths, $k$ is number of information qubits per block and $p$ is
 error probability of the channel.
\par
 For a fixed QCC, the distance of the code $d$ is finite and hence both $P_e$
 and $P_b$ scales only like a power law of $p$. Nevertheless, there are many
 $k$-input $n$-output QCCs with different memories $m$. And for many fixed $k$
 and $n$, the distance of the code $d$ increases approximately linearly with
 memory $m$. And such a family of codes may be constructed, for instance, from
 a corresponding family of classical convolutional codes. One such family of
 classical code as briefly discussed in Ref.~\cite{Convolution_Book} encodes
 one classical bit into two. By the construction of Chau in Ref.~\cite{QCC}, it
 can be turned into a family of 1-input 4-output QCCs. And as the memory $m$
 tends to infinity, both $P_e$ and $P_b$ becomes exponentially small. Thus, we
 have a family of good QCCs.
\section{Non-Catastrophic Quantum Codes}
\label{S:Good_QCC}
 The ability to decode a QCC is not sufficient to make QCC useful. We must also
 make sure that any decoding error will not propagate infinitely in spite of
 the convolutional nature of the code. To facilitate discussions, I borrow the
 following terminology from classical coding theory:
\begin{Def}
 A QCC is said to be {\bf catastrophic} if there exists a local decoding error
 that can propagate infinitely. Otherwise, a QCC is said to be
 {\bf non-catastrophic}. Clearly, useful QCCs must be non-catastrophic.
\end{Def}
\par\indent
 In case of classical convolutional codes and when the number of internal
 states per register $N$ is a prime power, a convolutional encoder can be
 mathematically represented by a polynomial of one variable over a finite
 field. Such a polynomial ring is clearly an Euclidean domain. In particular,
 two polynomials in an Euclidean domain have a unique greater common divisor
 (up to multiplication of units). Using this nice property of an Euclidean
 domain, Massey and Sain \cite{Inverse,Inverse_Pre} proved a necessary and
 sufficient condition for a classical convolutional code to be
 non-catastrophic.  Nonetheless, quantum mechanical operations are
 intrinsically non-commutative. Thus, the proof of Massey and Sain does not
 work for QCCs.
\par
 Quite surprisingly, a necessary and sufficient condition for a QCC to be
 non-catastrophic can still be found whenever $N$ is a prime power (and hence
 ${\Bbb Z}_N$ is a finite field). And I am going to report the criterion after
 introducing the following rather involved notations.
\subsection{Notation For Encoding And Decoding Qubits When $N$ Is A Prime
 Power}
\label{SS:Notation}
 In case $N$ is a prime power, any two body unitary operations can be generated
 by the span of the following elementary Pauli group operations
 \cite{Pauli_Group}: (a) addition $|x\rangle \mapsto |x\!+\!a\rangle$ and $|x,y
 \rangle \mapsto |x\!+\!y,y\rangle$ for some $a\in {\Bbb Z}_N$, (b)
 multiplication $|x\rangle \mapsto |ax\rangle$ for some $a\in {\Bbb Z}_N
 \backslash \{ 0\}$, (c) Fourier transform $|x\rangle \mapsto \sum_y
 \omega_N^{xy} |y\rangle$, (d) local phase multiplication $|x\rangle \mapsto
 \omega_N^{ax} |x\rangle$, and (e) non-local phase multiplication $|x,y\rangle
 \mapsto \omega_N^{xy} |x,y\rangle$.
\par
 Let me first considers those QCCs that can be encoded can be implemented by a
 $k$-input $n$-output $m$-memory quantum sequential circuit. In this case, I
 can group the initial unencoded quantum state and preset registers into blocks
 of length $n$. This state is spanned by $\{ \bigotimes_{i=1}^{+\infty}
 |x_{i1},x_{i2},\ldots ,x_{in}\rangle \}$, where $x_{ij}$ for $i\geq 1$ and $1
 \leq j\leq n\!-\!k$ are the preset registers and $x_{ij}$ for $i\geq 1$ and
 $n\!-\!k < j \leq n$ are the quantum information registers. With this notation
 in mind, I define the following operators: (a) {\em state delay operator}
 $D^m : |x_{i,j} \rangle \mapsto |x_{i-m,j}\rangle$, (b) {\em Fourier transform
 operator} $F_p : |x_{ij}\rangle \mapsto \sum_{x_{ij}} \omega_N^{x_{ij}
 y_{i,j-p}} |y_{i,j-p}\rangle$, (c) {\em phase projection operator}\footnote{I
 shall explain why I introduce such a non-invertible projection operator in the
 next paragraph.} $P : e^{i\phi} |x_{ij}\rangle \mapsto |x_{ij}\rangle$, (d)
 {\em local phase multiplication operator} $L^m : |x_{ij}\rangle \mapsto
 \omega_N^{m x_{ij}} |x_{ij}\rangle$, (e) {\em non-local phase multiplication
 operator} $M_{mp} : |x_{i,j}\rangle \mapsto \omega_N^{x_{i,j} x_{i-m,j-p}}
 |x_{i,j}\rangle$, (f) {\em state addition operator} $+ : ( e^{i\phi} |x
 \rangle, e^{i\phi'} |x'\rangle ) \mapsto e^{i (\phi+\phi')} |x+x'\rangle$, and
 (g) {\em state multiplication operator} $a : |x_{ij}\rangle \mapsto |ax_{ij}
 \rangle$. Using these notations, the five elementary Pauli group operations
 can be represented by (a) $1+D^m P$, (b) $a$, (c) $F_0$, (d) $L^a$, and (e)
 $M_{m0}$, respectively. More generally, if I write the initial state ket
 together with the preset and ancillary quantum registers in a $p\times 1$
 column vector, then the composition of several Pauli group operations can be
 represented by a $p\times p$ matrix whose elements belong to the
 non-commutative ring ${\Bbb Z}_N \left< D,M_{ij},P,L,F_i \right>$ with $F_i^2
 = -1$, $P^2 = P$, $PD = DP$, $L M_{ij} = M_{ij} L$, $PL = P$, $P M_{ij} = P$,
 $M_{ij} M_{pq} = M_{pq} M_{ij}$, and $M_{ij} m = m^{-1} M_{i,j}^m$ for $m\in
 {\Bbb Z}_N \backslash \{ 0 \}$.
\par
 Since the Pauli group spans the set of all two body operators
 \cite{Pauli_Group}, a general quantum encoding circuit $U_{\rm encode}$ for a
 $k$-input $n$-output $m$-memory QCC can be written as a finite sum $\sum_i
 (\alpha_i, g_i )$ where $\alpha_i \in {\Bbb C}$ and $g_i \in \left(
 {\Bbb Z}_N \left< D,M_{ij},P,L,F_i \right> \right)^{n+p,n+p}$ where $p$ is the
 number of ancillary quantum registers required in the encoding process per
 block. For instance, the operator $\sum_{i=0}^{N-1} (1/N, L^i )$ sends
 $|0\rangle$ to $|0\rangle$ and all other $|i\rangle$ to 0. Furthermore, the
 unitary operator sending $|x,y\rangle$ to $|x,x\!+\!y\rangle$ can be written
 as
\begin{equation}
 \left[ \begin{array}{cc} (1,1) & (0, 0) \\ (1,P) & (1,1) \end{array} \right]
 ~. \label{E:CNOT_Example}
\end{equation}
 Readers should observe that the phase projection operator $P$ in
 Eq.~(\ref{E:CNOT_Example}) is essential. If I replace $P$ by $1$ in
 Eq.~(\ref{E:CNOT_Example}), then the replaced operator will not be
 well-defined for it would have mapped $e^{i\theta_1 + \theta_2} |x,y\rangle$
 to $e^{i\theta_1 + \theta_2} |x,x\!+\!y\rangle$ and $e^{i\theta_1} |x\rangle
 \otimes e^{i\theta_2} |y\rangle$ to $e^{i(2\theta_1 + \theta_2}) |x,x\!+\!y
 \rangle$. In addition, the operator expressed in Eq.~(\ref{E:CNOT_Example}) is
 unitary in spite of its apparent non-skew symmetric form.
\par
 Let me denote the set of all finite sums in the form $\sum_i (\alpha_i, g_i )$
 by $K$. Then, if I forget about the initial preset and ancillary quantum
 registers and simply represent the initial unencoded quantum information as a
 $k\times 1$ column vector, then I can simply write a $k$-input $n$-output
 $m$-memory quantum encoding circuit as a $n\times k$ matrix in $K^{n,k}$. The
 decoding circuit for this QCC is equal to a $(n\!+\!p)\times (n\!+\!p)$ matrix
 $U_{\rm encode}^{-1}$. Nevertheless, $U_{\rm encode}^{-1} \not\in K^{n+p,n+p}$
 in general. Similar to the encoding circuit, if I forget about the initial
 preset and ancillary quantum registers used in the decoding circuit, then I
 can present the decoding circuit by a $k\times n$ matrix.
\begin{Eg}
 Using the above notation, the encoding and decoding algorithms for the
 classical non-catastrophic convolutional code (written in a quantum state ket
 form) $|k_1,k_2,\ldots\rangle \longmapsto \bigotimes_{i=1}^{+\infty}
 |k_i\!+\!k_{i-2},k_i\!+\!k_{i-1}\!+\!k_{i-2}\rangle$ can be written as
 \begin{equation}
  \left[ \begin{array}{c} (1, P[1+D^2]) \\ (1, D + P[1+D^2]) \end{array}
  \right] \label{E:Encode_CCC}
 \end{equation}
 and
 \begin{equation}
  \left[ \begin{array}{cc} (1, -D^{-1}) & (1, D^{-1}) \end{array} \right] ~,
  \label{E:Decode_CCC}
 \end{equation}
 respectively.
 \par
 Similarly, the encoding and decoding algorithms for the classical catastrophic
 convolutional code $|k_1,k_2,\ldots\rangle \longmapsto
 \bigotimes_{i=0}^{+\infty} |k_i\!+\!k_{i-1},k_i\!+\!k_{i-2}\rangle$ can be
 written as
 \begin{equation}
  \left[ \begin{array}{c} (1, P[1+D]) \\ (1, 1+P D^2) \end{array} \right]
  \label{E:Encode_CCC_Cata}
 \end{equation}
 and
 \begin{equation}
  \left[ \begin{array}{cc} (1, [1+PD]^{-1} D) & (1, [1+PD]^{-1}) \end{array}
  \right] ~, \label{E:Decode_CCC_Cata}
 \end{equation}
 respectively. \label{Eg:En_Classical_Notation}
\end{Eg}
\begin{Eg}
 One possible way to encode the quantum state $|k_1,k_2,\ldots\rangle$ as a
 QCC given by Eq.~(\ref{E:14QCC}) is as follows: First, I prepare a number of
 preset quantum registers and write the initial state as
 $\bigotimes_{i=1}^{+\infty} |k_i,0,0,0\rangle$. Then, I transform this state
 to $\bigotimes_{i=1}^{+\infty} |k_i,k_{i-1},k_{k-2},0\rangle$ by the unitary
 operator
 \begin{equation}
  \left[ \begin{array}{cccc} (1,1) & (0,0) & (0,0) & (0,0) \\ (1,P D) & (0,0) &
  (0,0) & (0,0) \\ (1,P D^2) & (0,0) & (1,1) & (0,0) \\ (0,0) & (0,0) & (0,0) &
  (1,1) \end{array} \right] ~. \label{E:En_1st}
 \end{equation}
 Then, I transform the state to $\bigotimes_{i=1}^{+\infty} |k_i\!+\!k_{i-2},
 k_i\!+\!k_{i-1}\!+\!k_{i-2},0,0\rangle$ by
 \begin{equation}
  \left[ \begin{array}{cccc} (1,1) & (0,0) & (1,P D) & (0,0) \\ (1, P D) & (1,
  1) & (1,P D^2) & (0,0) \\ (0,0) & (1, -P D) & (1,1) & (0,0) \\ (0,0) & (0,0)
  & (0,0) & (1,1) \end{array} \right] ~. \label{E:En_2nd}
 \end{equation}
 Next, I unitarily transform the state to $\bigotimes_{i=1}^{+\infty}
 \omega_N^{p_i (k_i+k_{i-2}) + q_i (k_i+k_{i-1}+k_{i-2})}$ $|p_i,q_i,
 0,0\rangle$ by
 \begin{equation}
  \left[ \begin{array}{cccc} (1,F_0) & (0,0) & (0,0) & (0,0) \\ (0,0) & (1,F_1)
  & (0,0) & (0,0) \\ (0,0) & (0,0) & (1,1) & (0,0) \\ (0,0) & (0,0) & (0,0) &
  (1,1) \end{array} \right] ~. \label{E:En_3rd}
 \end{equation}
 Finally, I bring the state to Eq.~(\ref{E:14QCC}) by the unitary
 transformation
 \begin{equation}
  \left[ \begin{array}{cccc} (1,P[1+D]) & (0,0) & (1,1) & (0,0) \\ (1,P[1+D]) &
  (1,PD) & (1,P) & (1,1) \\ (0,0) & (1,1) & (0,0) & (1,P) \\ (1,1) & (1,P) &
  (0,0) & (1,P) \end{array} \right] ~. \label{E:En_4th}
 \end{equation}
 Thus, the unitary encoding transformation $U_{\rm encode}$ for the QCC in
 Eq.~(\ref{E:14QCC}) simply equals to the product of the matrices in
 Eqs.~(\ref{E:En_1st})--(\ref{E:En_4th}). Moreover, if we forget about the
 initial preset registers, then the encoding operation is simply given by the
 first column of the matrix $U_{\rm encode}$, which is given by
 \begin{equation}
  \left[ \begin{array}{c} (1, P[1+D]F_0 [1+PD^2]) \\ (1, P[1+D]F_0 [1+PD^2] +
  PDF_1 [1+D+D^2]) \\ (1, [1+PD]F_1 P [1+D+D^2]) \\ (1, F_0 [1+PD^2] + P[1+D]
  F_1 P [1+D+D^2]) \end{array} \right] ~. \label{E:Encode_14QCC}
 \end{equation}
 Similarly, the decoding operation is equal to the first row of the matrix
 $U_{\rm encode}^{-1}$, namely,
 \begin{eqnarray}
  & & \left[ \begin{array}{c} (1, P[DF_1^{-1} - 1]) \\ (1, -PDF_1) \\ (1,[1+PD]
  F_0 P - PDF_1^{-1} - P[1+D]) \\ (1, [1+PD]F_0^{-1} + P[1+D]) \end{array}
  \right]^{\rm T} \nonumber \\ & = & \left[ \begin{array}{c} (1, P[-DF_1 (-1) -
  1]) \\ (1, -PDF_1) \\ (1,[1+PD]F_0 P + PDF_1 (-1) - P[1+D]) \\ (1, -[1+PD]F_0
  (-1) + P[1+D]) \end{array} \right]^{\rm T} ~. \label{E:Decode_14QCC}
 \end{eqnarray}
 \label{Eg:En_De_Notation}
\end{Eg}
\subsection{Criterion For Non-Catastrophic Quantum Code When $N$ Is A Prime
 Power}
\label{SS:Noncatastrophic}
 Now, let me report a useful lemma before proving a necessary and sufficient
 condition for non-catastrophic QCCs.
\begin{Lemma}
 Suppose $N$ is a prime power. And let $M\in K^{p,p}$ be a valid unitary
 operator acting on a possible infinitely long quantum signal. Then, $M$ can be
 decomposed into a product of finite product $\prod_{i=1}^q M_i$. Moreover, the
 $p^2$ elements in each matrix $M_i$ commute with each other for all $i$.
 \label{Lem:Matrix}
\end{Lemma}
\par\medskip\noindent
{\it Proof:} Since $N$ is a prime power and hence ${\Bbb Z}_N$ is a field, I
 can borrow the idea in Ref.~\cite{Decomposition} to decompose the matrix $M$
 as a product of finitely many matrices. Observe that I can always find an
 invertible $p\times p$ matrix $N_1$ such that the element located in the $p$th
 row and $(p\!-\!1)$th column in $N_1^{-1} M$ equals $(0,0)$. Besides, I can
 choose $N_1$ in such a way that the elements in the $i$th row and $j$th column
 satisfy the condition
\begin{equation}
 \left( N_1 \right)_{ij} = \left\{ \begin{array}{ll} (1,1) & \mbox{~~if~} i=j
 \mbox{~and~} i\leq p\!-\!2 \\ (0,0) & \mbox{~~if~} i\neq j \mbox{~and~} (i,j)
 \neq (p,p\!-\!1) \end{array} \right. ~.
 \label{E:Condition_N_1}
\end{equation}
 Similarly, I can find a $p\times p$ matrix $N_2$ such that $\left( N_2^{-1}
 N_1^{-1} M \right)_{ij} = (0,0)$ whenever $(i,j) = (p,p\!-\!1)$ and $(p,
 p\!-\!2)$. Moreover,
\begin{equation}
 \left( N_2 \right)_{ij} = \left\{ \begin{array}{ll} (1,1) & \mbox{~~if~} i=j
 \mbox{~and~} i\neq p\!-\!2 \mbox{~or~} p \\ (0,0) & \mbox{~~if~} i\neq j
 \mbox{~and~} (i,j) \neq (p,p\!-\!2) \end{array} \right. ~.
 \label{E:Condition_N_2}
\end{equation}
 Inductively, I can find $N_i$ such that $M' = N_{p(p-1)/2}^{-1}
 N_{[p(p-1)/2]-1}^{-1} \cdots N_1^{-1} M$ is a upper triangular matrix.
 Besides, at most three elements in $N_i\!-\!I_p$ are non-zero where $I_p$
 denotes the identity operator. Similarly, I can transform the matrix $M'$
 into a diagonal one by means of $p(p\!-\!1)/2$ matrices in a similar form as
 $N_i$. Thus, $M = N_1 N_2 \cdots N_{p(p-1)/2} M'' N_{[p(p-1)/2]+1}
 N_{[p(p-1)/2]+2} \cdots N_{p(p-1)}$ where $M''$ is a diagonal matrix and all
 $N_i\!-\!I_p$ can be brought into the form in Eq.~(\ref{E:Condition_N_1}) by
 relabeling some columns and rows plus possibly a transposition.
\par
 It is obvious that $M''$ is equal to a product of $p$ diagonal matrices, each
 of which has at most one diagonal element differ from $(1,1)$. Besides,
 elements in each of the $p$ matrices commute with each other. Thus, to
 complete the proof, it remains to show that $N_i$ can be decomposed into a
 finite product of matrices whose elements commute. In fact, it suffices for me
 to decompose it for the $2\times 2$ matrix $\tilde{M} = \left[
 \begin{array}{cc} A & B \\ (0,0) & C \end{array} \right]$. The decomposition
 for the matrix $N_i$ is similar. Since $\tilde{M}$ is a well-defined operator,
 either $A$ or $B$ (but not both) must be in the form $P X$ for some $X\in K$.
 In the first case, $A = PX$ and it is easy to check that
\begin{mathletters}
\begin{equation}
 \tilde{M} = \left[ \begin{array}{cc} (1,0) & (0,0) \\ (0,0) & C \end{array}
 \right] \,\left[ \begin{array}{cc} (1,P) & B \\ (0,0) & (1,0) \end{array}
 \right] \,\left[ \begin{array}{cc} X & (0,0) \\ (0,0) & (1,1) \end{array}
 \right] ~. \label{E:Decompose_M_T_1}
\end{equation}
 And in the second case, $B = PY$ and
\begin{equation}
 \tilde{M} = \left[ \begin{array}{cc} (1,0) & (0,0) \\ (0,0) & C \end{array}
 \right] \,\left[ \begin{array}{cc} (1,P) & Y \\ (0,0) & (1,0) \end{array}
 \right] \,\left[ \begin{array}{cc} X & (0,0) \\ (0,0) & (1,1) \end{array}
 \right] ~. \label{E:Decompose_M_T_2}
\end{equation}
\end{mathletters}
 Since elements in each of the matrices in the right hand sides of
 Eqs.~(\ref{E:Decompose_M_T_1}) and~(\ref{E:Decompose_M_T_2}) commute, so the
 lemma is proved.
\hfill$\Box$
\par\medskip
 After going through the above preparatory discussions and examples, I am ready
 to state a necessary and sufficient condition for non-catastrophic QCCs. In
 fact, Theorem~\ref{Thrm:Non_Cata} below generalizes a necessary and sufficient
 condition for classical non-catastrophic codes \cite{Inverse,Inverse_Pre}.
\begin{Thrm}
 The following statements concerning a $k$-input $n$-output $m$-memory QCC are
 equivalent when $N$ is a prime power:
\par
 (a) The QCC is non-catastrophic.
\par
 (b) There exits a quantum encoding circuit (that includes the preset and
 ancillary quantum registers) $g \in K^{n+p,n+p}$ for the QCC such that its
 left inverse $g^{-1}$ exists. Moreover, elements in the matrix $g^{-1}$ can be
 expressed as a finite sum $\sum_i (\alpha_i, g'_i)$ with $g'_i \in {\Bbb Z}_N
 \left< D,D^{-1},M_{pq},P,L,F_p \right>$ for all $i$.
\par
 (c) There exists a quantum encoding circuit that can be decomposed into the
 following finite product $g = \prod_i g_i$ in such a way that for each $i$,
 (1) $g_i \in K^{n+p,n+p}$, (2) elements of matrix $g_i$ belong to a
 commutative polynomial ring, and (3) the inverse $\left( \det g_i
 \right)^{-1}$ exists and can be expressed as a finite sum $\sum_i (\alpha_j,
 g'_j)$ with $g'_j \in {\Bbb Z}_N \left< D,D^{-1},M_{pq},P,L,F_p \right>$.
\par
 (d) The quantum encoding circuit (that excludes the preset registers) $h\in
 K^{n,k}$ can be expressed as a finite product of matrices $\prod_i h_i$ in
 such a way that each $i$, (1) $h_i \in K^{a_i,b_i}$, (2) elements of matrix
 $h_i$ belong to a commutative polynomial ring, and (3) the greatest common
 divisor $t_i$ of the determinant of all the $a_i \choose b_i$ sub-matrices of
 $h_i$ is invertible and $t_i^{-1} = \sum_j (\alpha_j, t'_{ij})$ with $t'_{ij}
 \in {\Bbb Z}_N \left< D,D^{-1},M_{pq},P,L,F_p \right>$. \label{Thrm:Non_Cata}
\end{Thrm}
\par\medskip\noindent
{\it Proof:} By suitably adding ancillary quantum registers as well as
 enlarging the encoding matrix to include those ancillary registers, it is easy
 to see that (d) $\Rightarrow$ (c) $\Rightarrow$ (b). Now, I move on to show
 that (b) $\Rightarrow$ (a). Since there is a decoding circuit that can be
 represented as a $(n+p)\times (n+p)$ matrix whose elements $\sum_i (\alpha_i,
 h_i )$ with $h_i \in {\Bbb Z}_N \left< D,D^{-1},M_{ij},P,L,F_i \right>$. In
 other words, decoding each quantum register in the code requires only
 information from a finite number of encoded quantum registers. Thus, if there
 are only a finite number of encoded quantum registers in error, then the
 decoding errors will only be localized in a finite number of quantum
 registers. Hence, the code is non-catastrophic.
\par
 To complete the proof, it remains for me to show that (a) $\Rightarrow$ (d).
 Recall that if $U_{\rm encode} \in K^{n,n}$ is the encoding circuit, then the
 decoding circuit equals $U_{\rm encode}^{-1}$. So, if statement (d) is false,
 then I can extend the $k\times n$ decoding circuit into a $n\times n$ one. And
 since $N$ is a prime power, so by Lemma~\ref{Lem:Matrix}, I can conclude that
 elements in the $k\times n$ decoding circuit are in the form $\sum_i
 (\alpha_i,h_i)$ where $h_i$ belongs to the formal power series non-commutative
 ring ${\Bbb Z}_N \left< \!\left< D,M_{ij},P,L,F_i \right> \!\right>$ but not
 every element in the decoding circuit belongs to ${\Bbb Z}_N \left< D,M_{ij},
 P,L,F_i \right>$. Consequently, there exists an encoded quantum register whose
 state affects the states of infinitely many decoded quantum registers. Thus,
 the QCC is catastrophic and this complete the proof.
\hfill$\Box$
\par\medskip
 Now, it is clear from the proof of Theorem~\ref{Thrm:Non_Cata} that if I first
 let the encoded quantum to go through the QVA and then I apply the unitary
 transformation $g^{-1}$ (which is the left inverse of $g$) to it, I can
 recover the original unencoded quantum information. In addition, it is also
 clear that any QCC that cannot be expressed as a $k$-input $n$-output
 $m$-memory sequential quantum circuit must be catastrophic. Moreover, the
 conclusion in Theorem~\ref{Thrm:Non_Cata} remains valid if I extend the
 meaning of $m$-memory QCC to include those QCCs whose encoding scheme depends
 on the state of a finite number of previous or future blocks.
\par
 One possible way to construct QCC is to start with a classical convolutional
 code $C$ \cite{QCC}. Chau showed that one can first encode a quantum signal
 using the classical code $C$, then one takes the local Fourier transform on
 each encoded quantum register, and finally one encodes the resultant state ket
 by the code $C$ again, one gets a QCC \cite{QCC}. Here, I show that the QCC
 generated this way inherits the error propagation behavior from its parent
 classical code.
\begin{Cor}
 Suppose $C$ is a $k$-input $n$-output classical convolutional code and $Q$ be
 the corresponding $k^2$-input $n^2$-output QCC obtained using the above
 method. Then $C$ is catastrophic if and only if $Q$ is catastrophic.
 \label{Cor:QCC}
\end{Cor}
\par\medskip\noindent
{\it Proof:} I write the quantum encoding scheme as a product of three matrices
 $g_1 g_2 g_3$ where $g_1$ and $g_3$ involve the symbols $D$ and $DP$ and $g_2$
 involves the symbol $F_i$. That is to say, $g_1$ and $g_3$ represent the
 initial and final encoding by the classical code $C$ and $g_2$ represents the
 local Fourier transform. Suppose $C$ is non-catastrophic, then clearly I can
 arrange $g_1$, $g_2$ and $g_3$ to satisfy statement~(d) in
 Theorem~\ref{Thrm:Non_Cata} \cite{Inverse,Inverse_Pre}. Hence, $Q$ is
 non-catastrophic. Conversely, if $C$ is a catastrophic code, then from the
 construction of $Q$, it is clear that one can always find a finite number of
 spin flip errors for $Q$ such that the decoding errors propagate infinitely.
 Hence $Q$ is catastrophic.
\hfill$\Box$
\par\medskip
 Corollary~\ref{Cor:QCC} implies that the QCC given in Eq.~(\ref{E:14QCC}) is
 non-catastrophic.
\section{Fault-Tolerant Computation Using Quantum Convolutional Codes}
\label{S:Fault_Tolerant}
 The ability to decode a non-catastrophic QCC is still not enough to make them
 truly useful. We have to impose the requirement that the QCC must have a
 fault-tolerant implementation so that quantum information processing can take
 place in the encoded form. In QBCs, we know that all stabilizer (block) codes
 have a fault-tolerant implementations
 \cite{Gott1,Kitaev,Aharonov,KLZ,P_Review,Gott2,Gott3,Steane} under suitable
 wiring of quantum gates. And now, I am going to generalize the theory of
 stabilizer code and its fault-tolerant implementations to the world of QCCs.
\par
 For stabilizer codes, I restrict myself to consider the case when $N = 2$.
 Recall that in the case of QBC and when $N=2$, if we denote the coding space
 of an $n$ qubit code by $T$, then the stabilizer of this code $S$ is some
 Abelian subgroup of the group ${\cal R}^{\otimes n}$ whose elements fixes $T$
 \cite{Cald3,Cald2,Gott1,Gott2,Gott3}. Besides, $S$ can be generated by a
 finite number of operations $g_i \in {\cal R}^{\otimes n}$, known as the
 generators of $S$. Finally, the encoded spin flip and phase change operations
 are specified in ${\cal R}^{\otimes n}$. These operations commute with the
 stabilizer $S$. More precisely, the codeword for a $k$-input $n$-output
 stabilizer QBC can be written (up to an overall normalization constant) as:
\begin{equation}
 |x_1,x_2,\ldots ,x_k\rangle \longmapsto \sum_{q_i=0}^1 \left[ \prod_j
 \overline{\sigma}_{x,j} \left( \prod_i g_i^{q_i} |0,0,\ldots ,0\rangle \right)
 \right] \label{E:stabilizer}
\end{equation}
 where $g_i$ and $\overline{\sigma}_{x,j}$ are the generators of the stabilizer
 and the encoded spin flip operation for $|x_j\rangle$, respectively
 \cite{Gott1}. In addition, the encoded phase shift operators
 $\overline{\sigma}_{z,j}$ exist in ${\cal R}^{\otimes n}$ for all $j$.
\par
 Generalizing the stabilizer (block) code formalism to the QCC world is easy.
 One only needs to be more careful in dealing with the infinite number of
 qubits and hence the infinite number of generators for the stabilizer. First,
 one replaces ${\cal R}^{\otimes n}$ by $\prod_{i=1}^{+\infty} {\cal R}$.
 Clearly, $\prod_{i=1}^{+\infty} {\cal R}$ and hence the stabilizer $S$ have a
 counterable number of generators. Thus, Eq.~(\ref{E:stabilizer}) holds for
 QCCs as $k\rightarrow\infty$. Besides, for a $m$-memory QCC, the encoded spin
 flip operators $\overline{\sigma}_{x,j}$ as well as the encoded phase shift
 operators $\overline{\sigma}_{z,j}$ act on no more than $\mbox{O} (n(m+1))$
 qubits. In this way, the fault-tolerant error syndrome measurement procedure
 in stabilizer block code \cite{Gott1,P_Review} directly applies to
 convolutional code. Finally, one concatenate the QCC with another stabilizer
 QBC to $L$ levels. Then, by correcting the errors in all levels concurrently,
 one achieves an error reduction from $\mbox{O} (\epsilon)$ to $\mbox{O}
 (\epsilon^L)$. Hence, Gottesman's \cite{Gott3} proof that all stabilizer codes
 have fault-tolerant implementation directly carries over to the QCC world.
 (See also Ref.~\cite{Steane} for related results.) Thus, non-catastrophic
 stabilizer QCC are good codes. Recently, Gottesman extended his theory to
 cover a large number of $N$ary fault-tolerant quantum codes using Pauli group
 \cite{Pauli_Group}. A direct consequence of his result is that we can easily
 construct many $N$ary fault-tolerant QCCs.
\par
 Finally, I go on the show that the QCC in Eq.~(\ref{E:14QCC}) is a stabilizer
 code. In fact, I prove something more general:
\begin{Thrm}
 Let $C$ be a classical convolutional code. And let $Q$ be the corresponding
 QCC as described in Corollary~\ref{Cor:QCC}. Then, $Q$ is a stabilizer code.
 \label{Thrm:stabilizer}
\end{Thrm}
\par\medskip\noindent
{\it Proof:} The proof follows directly from the three lemmas below.
\hfill$\Box$
\begin{Lemma}
 All classical binary (block or convolutional) codes are stabilizer codes.
 \label{Lem:Classical_stabilizer}
\end{Lemma}
\par\medskip\noindent
{\it Proof:} Without lost of generality, I consider a $m$-memory classical
 convolutional code. Then, the encoded spin flip operator
 $\overline{\sigma}_{x,j}$ is nothing but a finite number of $\sigma_x$ acting 
 the encoded qubits. Since the code is classical, the encoded state for each
 $|x_1,x_2,\ldots\rangle$ can simply be represented by a single state ket
 without any dummy summation index. More precisely, elements of the stabilizer
 are those commute with the encoded spin flip operators and at the same time
 can be expressed in the form $A_1 A_2 A_3 \cdots$ where $A_i$ acts on the
 $i$th encoded qubit and $A_i \in \{ \openone,\sigma_z \}$. Clearly, this kind
 of elements forms an Abelian subgroup of $\prod_{i=1}^{+\infty} {\cal R}$ and
 has a counterable number of generators. Hence, the lemma is proved.
\hfill$\Box$
\begin{Eg}
 The stabilizer associated with the classical block code $|k\rangle \longmapsto
 |kkk\rangle$ is generated by $\sigma_y\sigma_y\openone$ and $\sigma_y\openone
 \sigma_y$. Furthermore, the encoded spin flip and phase shift operators equal
 $\sigma_x\sigma_x\sigma_x$ and $\sigma_z\sigma_z\sigma_z$, respectively.
 \label{Eg:Classical_stabilizer}
\end{Eg}
\begin{Eg}
 The encoded spin flip operators for the classical convolutional code $|k_1,
 k_2,\ldots\rangle \longmapsto \bigotimes_{i=0}^{+\infty} |k_i\!+\!k_{i-2},
 k_i\!+\!k_{i-1}\!+\!k_{i-2}\rangle$ is given by $\sigma_x\sigma_x\openone
 \sigma_x\sigma_x\sigma_x\openone\openone\cdots$, $\openone\openone\sigma_x
 \sigma_x\openone\sigma_x\sigma_x\sigma_x\openone\openone\cdots$, $\openone
 \openone\openone\openone\sigma_x\sigma_x\openone\sigma_x\sigma_x\sigma_x
 \openone\openone\cdots$, and so on. The encoded phase shift operators are
 given by $\openone\openone\sigma_z\sigma_z\openone\openone\cdots$, $\openone
 \openone\openone\openone\sigma_z\sigma_z\openone\openone\cdots$, and so on.
 Besides, the stabilizer for this code is generated by $\sigma_z\sigma_z
 \openone\openone\cdots$, $\openone\sigma_z\openone\sigma_z\sigma_z\sigma_z
 \openone\openone\cdots$, $\sigma_z\sigma_z\openone\sigma_z\openone\sigma_z
 \sigma_z\sigma_z\openone\openone\cdots$, $\openone\openone\sigma_z\sigma_z
 \openone\sigma_z\openone\sigma_z\sigma_z\sigma_z\openone\openone\cdots$,
 $\openone\openone\openone\openone\sigma_z\sigma_z\openone\sigma_z\openone
 \sigma_z\sigma_z\sigma_z\openone\openone\cdots$, and so on.
 \label{Eg:Classical_Convolutional_stabilizer}
\end{Eg}
\begin{Lemma}
 Let $C$ be a classical binary code. And let $C'$ be the code obtained by
 locally Fourier transforming each qubits in the code $C$. Then, both $C$ and
 $C'$ are stabilizer codes. \label{Lem:FFT}
\end{Lemma}
\par\medskip\noindent
{\it Proof:} Lemma~\ref{Lem:Classical_stabilizer} says that $C$ is a stabilizer
 code. Suppose $g_i$ are the generators of the stabilizer of $C$,
 $\overline{\sigma}_{x,j}$ is the encoded spin flip operators of $C$ as
 described in Lemma~\ref{Lem:Classical_stabilizer}. Define $g_i'$ to be $g_i$
 with $\sigma_x$ replaced by $\sigma_z$. Similarly, I define
 $\overline{\sigma}'_{x,j}$ to be $\overline{\sigma}'_{x,j}$ but with
 $\sigma_x$ replaced by $\sigma_z$. Then, it is easy to verify that $g_i'$
 generate the stabilizer of $C'$. Besides, $\overline{\sigma}'_{x,j}$ is the
 encoded spin flip operator for the code $C'$.
\hfill$\Box$
\begin{Lemma}
 Let $C$ and $C'$ be the codes as described in
 Lemmas~\ref{Lem:Classical_stabilizer} and~\ref{Lem:FFT}. Then the code $C''$
 obtained by first encode the state by $C'$ and then by $C$ is a stabilizer
 code. (Compare with Ref.~\cite{Pasting} for a similar result.) In addition,
 each encoded spin flip and phase shift operator for $C''$ acts on a finite
 number of qubits provided that $C$ and hence $C''$ are non-catastrophic.
 \label{Lem:Composition}
\end{Lemma}
\par\medskip\noindent
{\it Proof:} Suppose $C$ and hence also $C'$ are $k$-input $n$-output codes
 with finite memory. Then from Lemmas~\ref{Lem:Classical_stabilizer}
 and~\ref{Lem:FFT}, I can write the generators of the stabilizer code $C$ as
 $A_{i1}A_{i2}A_{i3}\cdots$ where $A_{ij}\in\{ \openone,\sigma_x \}$. Moreover,
 I write $B_{i1}B_{i2}\cdots$ as the generators of the stabilizer code $C'$
 where $B_{ij}\in \{ \openone,\sigma_z \}$. Suppose $X_{i1}X_{i2}\cdots$ and
 $Z_{i1}Z_{i2}\cdots$ be the encoded spin flip operators for codes $C$ and
 $C'$, respectively. Recall that $C$ can be expressed in the form
 \cite{Convolution_Book}
\begin{equation}
 |x_1,x_2,\ldots\rangle \longmapsto \bigotimes_{i=1}^{+\infty} |\sum_j a_{ij}
 x_j\rangle ~. \label{E:Map1}
\end{equation}
 Since $C$ is of finite memory, the sum in each of the encode qubits in
 Eq.~(\ref{E:Map1}) above is finite. More precisely, $a_{ij} = 0$ or $1$ and
 for each fixed $i$, only a finite number of $a_{ij}$ equals one. Consequently,
 the QCC $C''$ can be expressed in the form
\begin{equation}
 |x_1,x_2,\ldots\rangle \longmapsto \bigotimes_{i=1}^{+\infty} \left[
 \sum_{p_1,p_2,\ldots} \omega_N^{\sum_j a_{ij} x_j p_i} |\sum_r b_{ir} p_r
 \rangle \right] ~, \label{E:Map2}
\end{equation}
 where $b_{ij} = 0$ or $1$ and for each fixed $i$, only a finite number of
 $b_{ij}$ equals one.
\par
 If $C$ is catastrophic, its decoding circuit can be expressed as a formal
 power series matrix. While if $C$ is non-catastrophic, its decoding circuit
 can be expressed as a polynomial matrix \cite{Inverse}. (See also
 Theorem~\ref{Thrm:Non_Cata}.) Consequently, there exist $c_{ij} \in \{ 0,1
 \}$ such that $\sigma_x^{c_{i1}}\sigma_x^{c_{i2}}\sigma_x^{c_{i3}}\cdots$
 is an operator acting on the codeword of $C''$ whose result is to map $p_i$ to
 $p_i\!+\!1$ for all $i$. Besides, $\sigma_z^{c_{i1}}\sigma_z^{c_{i2}}
 \sigma_z^{c_{i3}}\cdots$ is an operator acting on the codeword of $C''$ whose
 result is to multiply the codeword by a phase $(-1)^{p_i}$ for all $i$.
 Similarly, there exists $d_{ij} \in \{ 0,1 \}$ such that $\sigma_x^{d_{i1}}
 \sigma_x^{d_{i2}}\cdots$ is an operator acting on the codeword of $C$ whose
 result is to map $x_i$ to $x_i\!+\!1$, and that $\sigma_z^{d_{i1}}
 \sigma_z^{d_{i2}}\cdots$ is an operator acting on the codeword of $C$ whose
 result is to multiply the codeword by a phase $(-1)^{x_i}$ for all $i$.
 Furthermore, for each fixed $i$, only a finite number of $c_{ij}$ and $d_{ij}$
 equals one if $C$ is non-catastrophic.
\par
 Once I know how to add one to $x_i$ and $p_i$ as well as how to add phases
 $(-1)^{x_i}$ and $(-1)^{p_i}$ to the codewords of $C$ and $C'$ in the previous
 paragraph, I can use them to construct the encoded spin flip and phase shift
 operators for the code $C''$. They are given by
\begin{mathletters}
\begin{equation}
 \overline{\sigma}''_{x,i} = \sigma_z^{\sum_j d_{ij} c_{j1}}\sigma_z^{\sum_j
 d_{ij} c_{j2}}\sigma_z^{\sum_j d_{ij} c_{j3}}\cdots \label{E:sigma_C3_x}
\end{equation}
 and
\begin{equation}
 \overline{\sigma}''_{z,i} = \sigma_x^{\sum_j d_{ij} c_{j1}}\sigma_x^{\sum_j
 d_{ij} c_{j2}}\sigma_x^{\sum_j d_{ij} c_{j3}}\cdots ~, \label{E:sigma_C3_z}
\end{equation}
\end{mathletters}
 respectively.
\par
 After identifying the encoded spin flip and phase shift operations in $C''$,
 it remains for me to find the generators of the stabilizer of $C''$. First, by
 direct checking, I know that the operator $\sigma_x^{c_{i1}}\sigma_x^{c_{i2}}
 \cdots$ belongs to the stabilizer of $C''$ for all $i$. Then, similar to the
 proof of Lemma~\ref{Lem:Classical_stabilizer}, I consider operators in the form
 $A_{i1}A_{i2}\cdots$ with $A_{ij}\in \{\openone,\sigma_z\}$ that commute with
 the encoded spin flip, encoded phase shift and $\sigma_x^{c_{i1}}
 \sigma_x^{c_{i2}}\cdots$. Now, I choose a (counterable number of) generators
 amongst them. Then, the union of these operators and $\sigma_x^{c_{i1}}
 \sigma_x^{c_{i2}}\cdots$ generate with stabilizer of the code $C''$.
\hfill$\Box$
\begin{Eg}
 When $N=2$, the encoded spin flip operators for the QCC in Eq.~(\ref{E:14QCC})
 are $\sigma_z\sigma_z\sigma_z\sigma_z\sigma_x\sigma_x\openone\openone\sigma_z
 \sigma_z\sigma_z\sigma_z\openone\openone\cdots$, $\openone\openone\openone
 \openone\sigma_z\sigma_z\sigma_z\sigma_z\sigma_x\sigma_x\openone\openone$
 $\sigma_z\sigma_z\sigma_z\sigma_z\openone\openone\cdots$, and so on. In
 addition, the encoded phase shift operators are $\sigma_x\sigma_x\sigma_x
 \openone\sigma_x\openone\sigma_x\sigma_x\openone\openone\cdots$, $\openone
 \openone\openone\openone\sigma_x\sigma_x\sigma_x\openone\sigma_x\openone
 \sigma_x\sigma_x\openone\openone\cdots$, and so on. \label{Eg:stabilizer_14QCC}
\end{Eg}
\par\medskip\indent
 According to the proof of Lemma~\ref{Lem:Composition}, fault-tolerant
 computation is possible for all QCCs constructed using the method in
 Theorem~\ref{Thrm:stabilizer}. More importantly, if one starts with a
 non-catastrophic classical convolutional code $C$, then the fault-tolerant
 spin flip, phase shift, and controlled swapping for the QCC $C''$ constructed
 in Theorem~\ref{Thrm:stabilizer} can all be done in finite number of quantum
 gates. In fact, as long as I carefully wire my quantum circuit to prevent the
 spreading of quantum errors throughout all the qubits (see
 Ref.~\cite{P_Review} for the tips of how to do this), I can perform
 fault-tolerant quantum computation on this kind of QCCs. Suppose I have a
 quantum signal $|x_1,x_2,\ldots\rangle$ and if I follow the fault-tolerant
 computation wiring rule, I may even perform computation between the $i$th and
 $j$th encoded qubits in the above signal provided that their encoded spin flip
 and phase shift operators acts on distinct places in the encoded signal.
\section{Discussions}
\label{S:Discuss}
 In summary, I have generalized the VDA to QCC and have shown the optimality of
 QVA for a memoryless channel. In addition, I reported a simple way to test if
 a QCC is non-catastrophic. The key observation for all these is that a lot of
 the classical coding concepts can be ``quantized'' provided that one performs
 the relevant operations with care. Finally, I show that certain QCCs can
 perform fault-tolerant quantum computation. Since classical convolutional
 codes may be regarded as stabilizer codes and good classical convolutional
 codes exists, therefore I conclude that good QCCs and their decoding algorithm
 exist.
\acknowledgments
 I would like to thank Debbie Leung for her useful comments. This work is
 supported by the Hong Kong Government RGC grant HKU~7095/97P.

\end{multicols}

\begin{references}
\bibitem{9-bit} P. W. Shor, \pra {\bf 52}, 2493 (1995).
\bibitem{5-bit} R. Laflamme, C. Miquel, J. P. Paz and W. H. Zurek, \prl
 {\bf 77}, 198 (1996).
\bibitem{St1} A. M. Steane, \pra {\bf 54}, 4741 (1996).
\bibitem{Cald3} A. R. Calderbank, E. M. Rains, P. W. Shor, and N. J. A. Sloane,
 IEEE\ Trans.\ Inf.\ Theo. {\bf 44}, 1369 (1998).
\bibitem{Cald2} A. R. Calderbank, E. M. Rains, P. W. Shor, and N. J. A. Sloane,
 \prl {\bf 78}, 405 (1997).
\bibitem{Gott1} D. Gottesman, {\it Stabilizer Codes and Quantum Error
 Correction} (Ph.D. Thesis, Caltech, 1997).
\bibitem{9-reg} H. F. Chau, \pra {\bf 55}, 839 (1997).
\bibitem{5-reg} H. F. Chau, \pra {\bf 56}, 1 (1997).
\bibitem{N-add} E. M. Rains, R. H. Hardin, P. W. Shor and N. J. A. Sloane,
 \prl {\bf 79}, 953 (1997).
\bibitem{AnE1} S. L. Braunstein, \prl {\bf 80}, 4084 (1998).
\bibitem{AnE2} S. Lloyd, and J.-J. E. Slotine, \prl {\bf 80}, 4088 (1998).
\bibitem{Cond2} E. Knill, Los Alamos preprint archive {\tt quant-ph/9608048}
 (1996).
\bibitem{QCC} H. F. Chau, \pra {\bf 58}, 905 (1998).
\bibitem{Viterbi} A. J. Viterbi, IEEE\ Trans.\ Inf.\ Theo. {\bf 13}, 260
 (1967).
\bibitem{Viterbi_Review} A. J. Viterbi, IEEE\ Trans.\ Comm.\ Tech. {\bf 19},
 751 (1971).
\bibitem{Forney} G. D. Forney, Jr., IEEE\ Trans.\ Inf.\ Theo. {\bf 16}, 720
 (1970).
\bibitem{Convolution_Book} A. Dholakia, {\it Introduction to convolutional
 codes with applications} (Kluwer, Boston, 1994), chap.~5.
\bibitem{Omura} J. K. Omura, IEEE\ Trans.\ Inf.\ Theo. {\bf 15}, 177 (1969).
\bibitem{Forney2} G. D. Forney, Jr., Inf.\ \&\ Control {\bf 25}, 222 (1974).
\bibitem{Viterbi2} A. J. Viterbi, IEEE\ Trans.\ Inf.\ Theo. {\bf 13}, 260
 (1967).
\bibitem{Fano} R. M. Fano, IEEE\ Tran.\ Inf.\ Theo. {\bf 9}, 64 (1963).
\bibitem{QKD_Sec} H.-K. Lo, and H. F. Chau, Science {\bf 283}, 2050 (1999) and
 the associated supplementary materials available at
 http://www.sciencemag.org/feature/data/984035.shl.
\bibitem{Pauli_Group} D. Gottesman, {\it Proc. 1st NASA Int. Conf. Quant.
 Comm. \& Quant. Comp.}, to appear (1998); also available on Los Alamos
 preprint archive {\tt quant-ph/9802007}.
\bibitem{Decomposition} M. Reck, A. Zeilinger, H. J. Bernstein and P. Bertani,
 \prl {\bf 73}, 58 (1994).
\bibitem{Inverse} J. L. Massey, and M. K. Sain, IEEE\ Trans.\ Comp. {\bf 17},
 330 (1968).
\bibitem{Inverse_Pre} J. L. Massey, and M. K. Sain,
 IEEE\ Trans.\ Auto.\ Control {\bf 12}, 644 (1967).
\bibitem{Kitaev} A. Yu. Kitaev, Russ.\ Math.\ Surv. {\bf 52}, 1191 (1997).
\bibitem{Aharonov} D. Aharonov, and M. Ben-Or, in {\it Proc. 29th Annual ACM
 Sym. on the Theo. of Comp.}, (ACM, New York, 1998), p.~176.
\bibitem{KLZ} E. Knill, R. Laflamme and W. Zurek, Science {\bf 279}, 342
 (1998).
\bibitem{P_Review} J. Preskill, in {\it Introduction to Quantum Computation and
 Information}, eds. H.-K. Lo, T. Spiller and S. Popescu (World Scientific,
 Singapore, 1998), p.~213.
\bibitem{Gott2} D. Gottesman, \pra {\bf 54}, 1862 (1996).
\bibitem{Gott3} D. Gottesman, \pra {\bf 57}, 127 (1998).
\bibitem{Steane} A. M. Steane, Los Alamos preprint archive
 {\tt quant-ph/9809054} (1998).
\bibitem{Pasting} D. Gottesman, Los Alamos preprint archive
 {\tt quant-ph/9607027} (1996).
\end{references}
\end{document}